\documentclass[sn-mathphys]{sn-jnl}

\usepackage{physics}
\usepackage{graphicx}
\usepackage{caption}
\usepackage{tablefootnote}
\usepackage{multicol}
\usepackage{amsmath}
\usepackage{multirow}
\usepackage{amssymb}
\usepackage{pifont}
\usepackage{amsfonts}
\usepackage{tabularx}
\usepackage{multirow}
\usepackage{bbold}
\jyear{2021}%
\theoremstyle{thmstyleone}%

\theoremstyle{thmstyletwo}%
\usepackage{tabularx}
\usepackage{multirow}
\usepackage{tabularx,multirow,array}
\usepackage{float}
\theoremstyle{thmstylethree}%

\begin{document}
\title[Quantum and semi--quantum key distribution in networks]{Quantum and semi--quantum key distribution in networks}
\author*[]{\fnm{Rajni} \sur{Bala}}\email{Rajni.Bala@physics.iitd.ac.in}
\author[]{\fnm{Sooryansh} \sur{Asthana}}\email{sooryansh.asthana@physics.iitd.ac.in}
\author[]{\fnm{ V.} \sur{Ravishankar}}\email{vravi@physics.iitd.ac.in}

\affil*[]{\orgdiv{Department of Physics}, \orgname{IIT Delhi}, \orgaddress{\street{Hauz Khas}, \city{New Delhi}, \postcode{110016}, \state{Delhi}, \country{India}}}

\abstract{ In this paper, we utilize the potential offered by \textit{multidimensional separable states} (MSS) for secure and \textit{simultaneous} distributions of keys in a \textit{layered network}. We present protocols for both \textit{quantum and semi-quantum} key distribution and discuss their robustness against various eavesdropping strategies. We provide a procedure to identify the requisite resource states  to generalize these protocols for arbitrary layered networks. Finally, we study the interrelation between the local dimensionalities of states and achievable key rates in a given layer. These proposals are realizable with current technology, thanks to the employment of MSS and many advances in the generation, manipulation, and measurement of higher-dimensional  orbital angular momentum states of light.

}


\maketitle


\abstract{

}

\maketitle
\section{Introduction}
\label{Introduction}
\begin{figure}[h]
    \centering
    \includegraphics[width=0.65\textwidth]{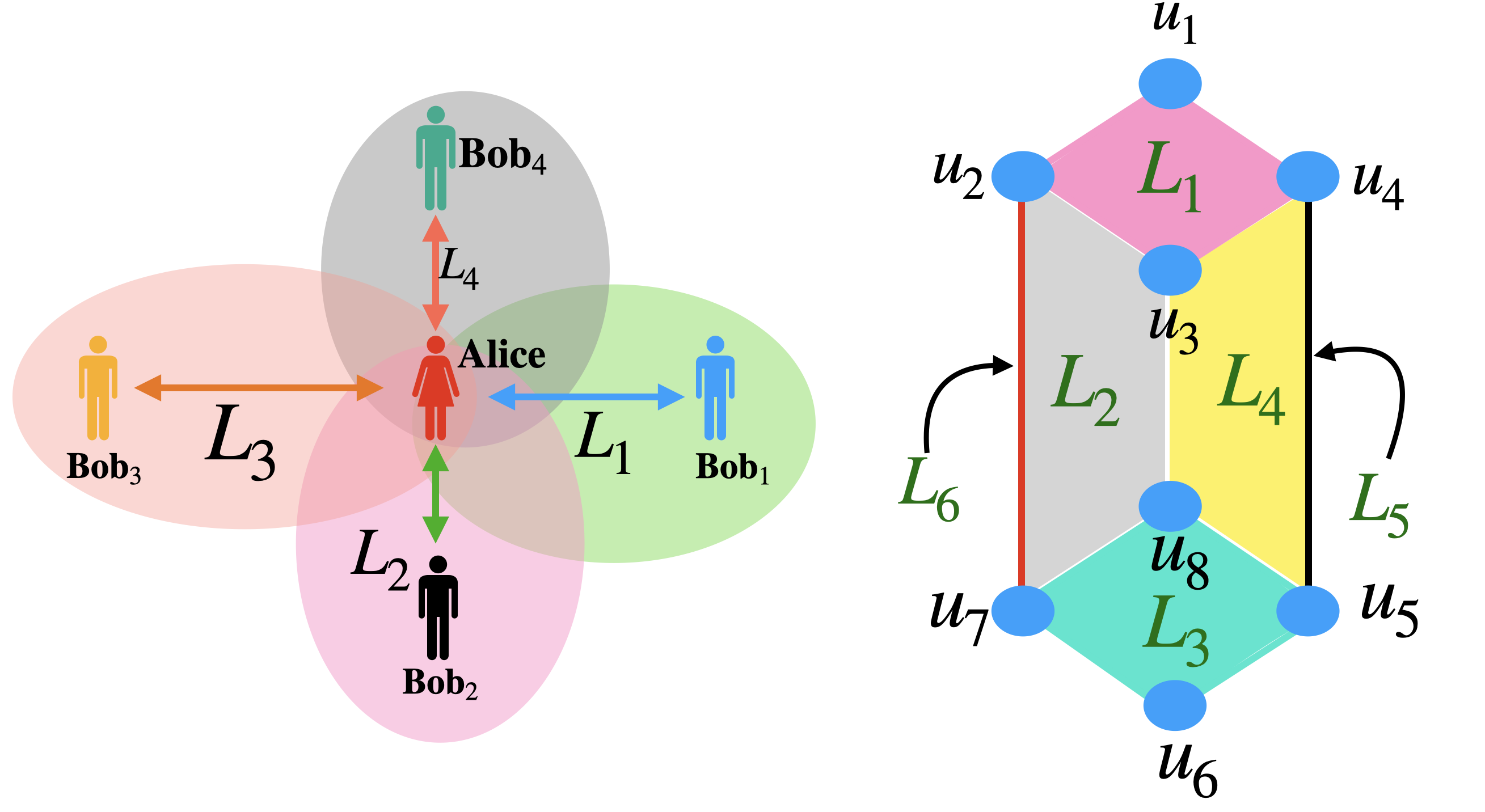}
    \caption{Pictorial illustrations of quantum networks :  (left) a star network in which Alice shares one key each with four participants namely Bob$_1$, Bob$_2$, Bob$_3$, and Bob$_4$. (right) A network in which keys are to be shared in distinct layers $L_i$. A participant may belong to more than one layer.  $L_1$ and $L_3$ consist of participants $\{u_1, \cdots, u_4\}$ and $\{u_5,\cdots, u_8\}$ respectively. $L_2$ and $L_4$ consist of $\{u_2, u_3, u_7, u_8\}$ and $\{u_3, u_4, u_5, u_8\}$ respectively. $L_5$ and  $L_6$ consist of $\{u_4, u_5\}$ and $\{u_2, u_7\}$ respectively. }
    \label{quantum network}
\end{figure}

The field of quantum key distribution (QKD) has witnessed vibrant research activity since the proposal of BB84 protocol   \cite{Bennett84,Ekert91,gisin2002quantum, Bechmann00,Bala2021}. There has been a two-pronged approach-- in the first approach, the focus has been on proposing protocols robust against side-channel attacks \cite{vazirani2019fully,lo2012measurement} and the other approach focused on sharing a secure key with  minimal resources. The latter approach has given rise to a semi-QKD (SQKD) protocol in which it suffices to have only one quantum participant\footnote{ By a quantum participant, one means a participant who can prepare any state and perform a measurement in any basis. In contrast, a classical participant can prepare a state and perform a measurement \textit{only in the computational basis} \cite{Boyer07}.  } \cite{Boyer07}. The large research interest in this idea owes to the flexibility offered to the participants in having access to the available classical resources \cite{yan2019semi, iqbal2020semi,li2020new}. These semi-quantum protocols harness the developed infrastructure of classical communication and use quantum resources in their minimal form to share secure information. This particular idea reduces the burden on the capabilities of resources at each participant's end and hence provides an edge to distribute information in a secure manner with only limited resources.  
 Subsequently, the study of the distribution of quantum information over a network, owing to  a plethora of possibilities that arise in realistic scenarios, has become a thriving field \cite{simon2017towards,cavalcanti2015detection,huang2009controlled,shukla2021hierarchical,epping2017multi} (see figure (\ref{quantum network}) for a pictorial representation of quantum networks).  For example, there may be a demand for secure quantum communication among different subsets of participants. Each of these subsets constitutes 
 what we call a {\it layer}. In particular, a protocol has been proposed for the secure distribution of keys in a layered network with all the quantum participants \cite{pivoluska2018layered}. 
 
 Elegant though the protocols proposed in \cite{pivoluska2018layered} are, they nevertheless require multidimensional entanglement as a resource. Though there is some experimental development in the generation of multidimensional entangled OAM states \cite{malik2016multi}, it is yet in its infancy. For example, three-photon asymmetric entangled states with Schmidt vectors $(3,3,2)$ and $(4,4,2)$ have been prepared with fidelities of $80.1\%$ and $85.4\%$, but with very small count rates of roughly $15~{\rm mHz}$ and $0.66~{\rm Hz}$ respectively \cite{erhard2020advances,hu2020experimental}. These states serve the purpose of  distributing secure keys in small networks having three participants and two layers. However, for large networks (which appear in realistic situations), the requisite layered entangled states have not yet been experimentally realized. Additionally, the yield of generation of such states puts another constraint as their generation involves non-linear processes \cite{malik2016multi}. Due to these limitations, these protocols are termed next-generation protocols \cite{pivoluska2018layered}.

These issues have motivated us to propose protocols involving resource states which can be generated even for large networks  with current technology. There has been a great advancement in the generation, manipulation, and measurement of higher-dimensional states such as orbital angular momentum (OAM) modes of light \cite{willner2021orbital}. Apart from the generation of OAM states with definite polarisation,  higher dimensional separable states have also been produced \cite{PhysRevApplied.11.064058}. In fact, weak coherent pulses (WCPs) of light carrying OAM have been generated with a relatively higher count rate of $4~{\rm kHz}$ \cite{mirhosseini2015high}. All these developments suggest that separable states of light can be used for communication in large networks. 

Therefore, we employ multidimensional separable states (MSS) as resources to distribute simultaneous keys in a layered network. We first develop a prepare-and-measure (P $\&$ M) layered QKD protocol. The protocol distributes keys in all the layers of a network having quantum participants in one go. As a further improvement, we develop a second protocol, {\it viz.}, P $\&$ M layered SQKD. The protocol requires only one participant to have access to quantum resources. Like other semi-quantum protocols, this protocol takes its motivation from the costly nature of quantum resources. Hence, it reduces their usage by allowing all but one participant to have access only to classical resources. Thus, the protocol allows for the 
simultaneous distribution of keys in a network with minimal available resources, i.e., well-developed infrastructure of classical communication, and uses quantum resources in their minimal form to share secure information.

We start with the example of a network having two layers and three participants (sections (\ref{Illustrative_QKD}) and (\ref{SQKD})) and then evolve a procedure that can be generalized to arbitrarily structured networks (section (\ref{generalisation})).  The examples capture all the essentials of the protocols.
We discuss the robustness of the protocols against various eavesdropping strategies (sections (\ref{security}) and (\ref{security1})). The layered structure of the network allows sharing of keys in some of the layers even if there is eavesdropping in others. This would require pinpointing the layers in which there is eavesdropping. We show this is possible by pinpointing the location of the eavesdropper (section (\ref{pinpoint})).   
For a quick comparison, we compare various features of both layered QKD and layered SQKD with those of already existing ones, in the tables (\ref{tab:QKD}) and (\ref{tab:SQKD}) respectively. 
Finally, we study the interrelation in the dimensionality of subsystems and achievable key rate in a given layer (section (\ref{Tradeoff})). Section (\ref{conclusion}) concludes the paper.


  \begin{table}[h!]
 \begin{center}
 \resizebox{!}{3.25cm} {
    \centering{
\begin{tabular}{ | c| c|c|}
\hline
\multirow{3}{*}{}&\multirow{3}{2.25cm}{\centering \textbf{ Entanglement based layered QKD \cite{pivoluska2018layered}}} &\multirow{3}{2.5cm}{\bf Proposed P $\&$ M Layered QKD}\\
 && \\
&&\\\hline
\multirow{3}{*}{\centering Quantum channel} &\multirow{3}{*}{\centering ideal}&\multirow{3}{*}{\centering ideal}\\
&&\\
&&\\\hline
\multirow{3}{*}{\centering Resource states} &\multirow{3}{2.5cm}{\centering multidimensional entangled states}&\multirow{3}{2.5cm}{\centering multidimensional separable states}\\
&&\\
&&\\\hline
 \multirow{3}{3cm}{\centering  yield of resource states}&\multirow{3}{*}{\centering $\sim mHz$ \cite{erhard2020advances} }&\multirow{3}{*}{\centering $\sim kHz$ \cite{mirhosseini2015high}}\\
 &&\\
&&\\\hline
\multirow{3}{*}{\centering Current state-of-the-art for a network}&\multirow{3}{*}{\centering network of three participants}&\multirow{3}{*}{\centering in large networks}\\
 &&\\
&&\\ \hline
\multirow{3}{*}{\centering $\#$ Classical participants}&\multirow{3}{*}{\centering none}&\multirow{3}{*}{\centering none}\\
 &&\\
&&\\ \hline
\multirow{3}{*}{\centering Communication}&\multirow{3}{*}{\centering one-way}&\multirow{3}{*}{\centering one-way}\\
 &&\\
&&\\ \hline
\multirow{3}{3cm}{\centering $\#$ bases used for key generation}&\multirow{3}{*}{\centering $1$}&\multirow{3}{*}{\centering $2$}\\
 &&\\
&&\\ \hline
\multirow{3}{*}{\centering Sifted key rate}&\multirow{3}{*}{\centering $1$ bit per transmission}&\multirow{3}{*}{\centering $1$ bit per transmission}\\
 &&\\
&&\\ \hline
\multirow{3}{2.5cm}{\centering Scope for generalised network}&\multirow{3}{*}{\centering limited}&\multirow{3}{*}{\centering relatively higher}\\
 &&\\
&&\\ \hline
 \end{tabular}}
 \vspace{0.3cm}
      }
    \end{center}
    \caption{Comparison of the P$\&$M layered QKD protocol proposed in this paper with the protocol proposed in \cite{pivoluska2018layered}.}
      \label{tab:QKD}
\end{table}

\begin{table}[h!]
 \begin{center}
 \resizebox{!}{3.25cm} {
    \centering{
\begin{tabular}{ | c| c| c |c|}

\hline
\multirow{3}{*}{}&\multirow{3}{2.25cm}{\centering \textbf{  Multi-party SQKD \cite{zhou2019multi}}}&\multirow{3}{1.75cm}{\centering\bf Layered SQKD \cite{bala2022layered}} &\multirow{3}{3cm}{\bf Proposed P $\&$ M Layered SQKD }\\
 &&& \\
&&&\\\hline
\multirow{3}{*}{\centering Quantum channel} &\multirow{3}{*}{\centering ideal}&\multirow{3}{*}{\centering ideal}&\multirow{3}{*}{\centering ideal}\\
&&&\\
&&&\\\hline
\multirow{3}{*}{\centering Resource states} &\multirow{3}{2.5cm}{\centering cluster state}&\multirow{3}{2.5cm}{\centering multidimensional entangled states}&\multirow{3}{2.5cm}{\centering multidimensional separable states}\\
&&&\\
&&&\\\hline
 \multirow{3}{3cm}{\centering  yield of resource states}&\multirow{3}{*}{\centering $\sim mHz$ \cite{kiesel2005experimental}}& \multirow{3}{*}{\centering $\sim mHz$ \cite{erhard2020advances}}&\multirow{3}{*}{\centering $\sim kHz$ \cite{mirhosseini2015high}}\\
 &&&\\
&&&\\\hline
\multirow{3}{*}{\centering $\#$ Quantum participants}&\multirow{3}{*}{\centering only one}&\multirow{3}{*}{\centering only one}&\multirow{3}{*}{\centering only one}\\
 &&&\\
&&&\\ \hline
\multirow{3}{*}{\centering $\#$ layers}&\multirow{3}{*}{\centering single layer}&\multirow{3}{*}{\centering multi-layer}&\multirow{3}{*}{\centering multi-layer}\\
 &&&\\
&&&\\ \hline
\multirow{3}{*}{\centering Communication}&\multirow{3}{*}{\centering two-way}&\multirow{3}{*}{\centering two-way}&\multirow{3}{*}{\centering two-way}\\
 &&&\\
&&&\\ \hline
\multirow{3}{3cm}{\centering $\#$ bases used for key generation}&\multirow{3}{*}{\centering $1$}&\multirow{3}{*}{\centering $1$}&\multirow{3}{*}{\centering $1$}\\
 &&&\\
&&&\\ \hline
\multirow{3}{*}{\centering Sifted key rate}&\multirow{3}{*}{\centering $1$ bit per transmission}&\multirow{3}{*}{\centering $1$ bit per transmission}&\multirow{3}{*}{\centering $1$ bit per transmission}\\
 &&&\\
&&&\\ \hline
\multirow{3}{2.5cm}{\centering Scope for generalised network}&\multirow{3}{*}{\centering limited}&\multirow{3}{*}{\centering limited}&\multirow{3}{*}{\centering relatively higher}\\
 &&&\\
&&&\\ \hline
 \end{tabular}}
 \vspace{0.3cm}
       }
    \end{center}
    \caption{Comparison of the proposed P$\&$M layered SQKD protocol in this paper with the protocols proposed in \cite{zhou2019multi} and \cite{bala2022layered}.}
 \label{tab:SQKD}
\end{table}

\section{Prepare-and-measure QKD protocol in layered networks}
\label{PandMlayeredQKD}
In this section, we present a protocol that allows transferring keys securely in distinct layers in \textit{one go}, alleviating the need for running multiple parallel QKD protocols. We start with an example and then provide a procedure to generalize it to  an arbitrarily structured layered network.

\subsection{Illustrative protocol}
\label{Illustrative_QKD}
We consider a simple network of three participants (Alice, Bob$_1$, and Bob$_2$) and two layers ($L_1$ and $L_2$).  Layer $L_1$ consists of Alice and Bob$_1$ and layer $L_2$ consists  of Alice, Bob$_1$, and Bob$_2$. As in other QKD protocols, Alice randomly prepares states from two bases and sends them to Bob$_1$ and Bob$_2$ who measure in any of them.\\ 

\noindent\textbf{\textit{Aim:}} Simultaneous distribution of keys in layers $L_1$ and $L_2$. \\
\noindent\textbf{\textit{Resources:} } To implement the task, Alice employs  the following \textit{$4\otimes 2$ dimensional separable} states\footnote{Alice sends a four-dimensional subsystem to Bob$_1$ and a two-dimensional subsystem to Bob$_2$.  The choice of this local dimensionality for each state is dictated by the number of layers to which each participant belongs.},
\begin{align}\label{eq:sets}
    S_1: \Big\{\ket{00}, \ket{11}, \ket{20}, \ket{31}\Big\},~ ~S_2: \Big\{\ket{0'+}, \ket{1'-}, \ket{2'+}, \ket{3'-}\Big\},~
   \end{align}
   as a resource. The states in the set $S_2$ are Fourier transforms of the states in $S_1$ and are defined as:
   \begin{align}
   \label{Both_bases }
  & \ket{0'}\equiv \frac{1}{2}\big( \ket{0}+\ket{1}+\ket{2}+\ket{3}\big),~\ket{1'}\equiv \frac{1}{2}\big( \ket{0}-\ket{1}+\ket{2}-\ket{3}\big),\nonumber\\
  &\ket{2'}\equiv \frac{1}{2}\big( \ket{0}+\ket{1}-\ket{2}-\ket{3}\big),~\ket{3'}\equiv \frac{1}{2}\big( \ket{0}-\ket{1}-\ket{2}+\ket{3}\big),~\nonumber\\
  &~\ket{\pm} = \frac{1}{\sqrt{2}}\big(\ket{0}\pm \ket{1}\big).
   \end{align} 
   
\begin{figure}[h]
    \centering
    \includegraphics[width=11.5cm]{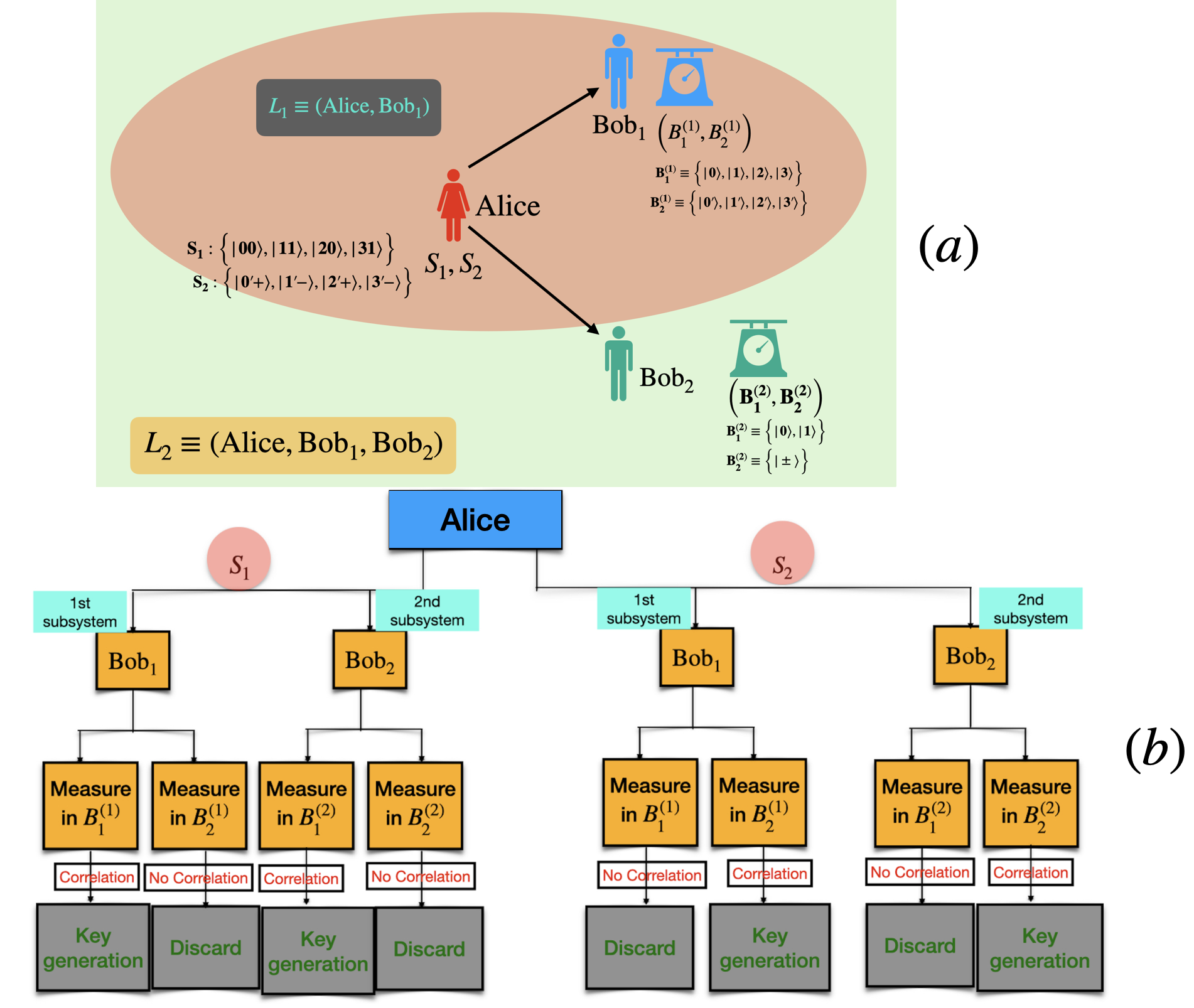}
    \caption{Schematic diagram of the  layered QKD protocol in a network. Figure (a) provides a description of the network and  required resources. Figure (b) provides a description of the protocol.}
    \label{fig:QKD1}
\end{figure}

For measurements on the received states, Bob$_1$ and Bob$_2$ employ the following sets of bases:
\begin{align}
    B_1^{(1)} &\equiv \{\ket{0}, \ket{1}, \ket{2}, \ket{3}\},~~~B_2^{(1)} \equiv \{\ket{0'}, \ket{1'}, \ket{2'}, \ket{3'}\},\nonumber\\
    B_1^{(2)} &\equiv \{\ket{0}, \ket{1}\};~~~~B_2^{(2)} \equiv \{\ket{+}, \ket{-}\},
    \end{align}
    where the superscripts  refer to the appropriate Bob. 
Note that the bases $B_1^{(1)}$ and $B_1^{(2)}$ are the computational bases and the bases $B_2^{(1)}$ and $B_2^{(2)}$ are their Fourier transformed bases.\\
The structure of the network and required resources together with the description of the protocol have been depicted in the figure (\ref{fig:QKD1}).
    \subsection*{\bf The protocol}
The steps of the protocol are as follows:
\begin{enumerate}
    \item Alice prepares a state randomly  from either the set $S_1$ or $S_2$, chosen with an equal probability. She records the quantum numbers of the states  with herself and sends the first and the second subsystems to Bob$_1$ and Bob$_2$ respectively.
    \item Bob$_1$ measures the received states in one of the bases $B^{(1)}_1$ and $B^{(1)}_2$, again chosen with an equal probability. Similarly, Bob$_2$ measures the received states in one of the bases $B^{(2)}_1$ and $B^{(2)}_2$  with an equal probability.
    \item This process is repeated for a large number of rounds.
    \item The outcomes of the rounds, in which Alice and Bobs have made any of the following choices, are kept:
   \begin{center}
          \resizebox{!}{0.7cm}{
\begin{tabular}{ |c|c|c| } 
 \hline
 Alice's set & Bob$_1$'s measurement basis & Bob$_2$'s measurement basis \\ 
 \hline
 $S_1$ & $B_1^{(1)}$ & $B_1^{(2)}$ \\ 
  \hline
  $S_2$& $B_2^{(1)}$ & $B_2^{(2)}$\\ 
 \hline
\end{tabular}}\end{center} 
Irrespective of Bob$_2$'s measurement, the rounds in which Alice and Bob$_1$'s measurement satisfies the above choice also  constitute key symbols in the layer $L_1$.
The data of  the rest of the rounds are discarded.
    \item {\it Check for eavesdropping:} Alice, Bob$_1$ and Bob$_2$ choose a subset of rounds to check for eavesdropping. If there is no eavesdropping, the outcomes of Alice and both the Bobs match, otherwise the protocol is aborted. 
    \item {\it Key generation:} Each participant expresses his/her outcomes in binary representation, i.e., $o=2o_1+o_0\equiv (o_1o_0)$. The symbols at the unit's place constitute a key in layer $L_2$ and those at the two's place correspond to key symbols in layer $L_1$. The same has been shown in the table (\ref{tab:key_rule_QKD}). 
    \end{enumerate}
\begin{table}[h!]
 \centering
 \resizebox{!}{1cm} {
\begin{tabular}{ |c|c|c|c| } 
 \hline
\multirow{2}{3cm}{Bob$_1$'s outcome in binary representation} & \multirow{2}{2cm}{Bob$_2$'s outcome} & \multirow{2}{2cm}{Key symbol in Layer $L_1$}&\multirow{2}{2cm}{ Key symbol in Layer $L_2$ }\\ 
&&&\\
 \hline
 $0\equiv (00)$ & $0$ & $0$&$0$ \\ 
  \hline
  $2\equiv (10)$& $0$ & $1$&$0$\\ 
 \hline
  $1\equiv (01)$& $1$ & $0$&$1$\\ 
 \hline
  $3\equiv (11)$& $1$ & $1$&$1$\\ 
 \hline
\end{tabular}}
\caption{Key generation rule}
\label{tab:key_rule_QKD}
\end{table}
In both the layers, since binary symbols are generated with equal probabilities, the sifted key rate in both the layers is $1$ bit each\footnote{Raw key rate in layer $L_2$ will have an additional factor of $1/2$ due to an extra participant in comparison to layer $L_1$.}.

\subsubsection{Confidentiality of keys} \label{confidentiality}
Another important consideration in the simultaneous sharing of keys in different layers is the confidentiality of keys. Thanks to  the choices of the resource states, the shared keys are completely confidential as  explained below.

In the network, Bob$_2$ is the only participant who does not belong to layer $L_1$. Therefore, certifying the confidentiality of keys is equivalent to  certifying that Bob$_2$ cannot get any information about the key shared in layer $L_1$. The choice of resource states in equation (\ref{eq:sets}) ensures that for respective outcomes of Bob$_2$ \textit{viz.} $0$ or $1$, Bob$_1$ gets two outcomes $0/2$ or $1/3$ randomly with equal probabilities. Thanks to complete randomness in the outcomes, Bob$_2$ cannot guess whether the key symbol being shared in layer $L_1$ is $0$ or $1$ (please see table (\ref{tab:key_rule_QKD})) and hence, the shared keys remain confidential.  This example makes the appropriateness of the choice explicit.
\subsection{Generalization to an arbitrary layered network}\label{generalisation}

 In this section, we  generalize  the protocol presented in the previous section to an arbitrarily  layered network.  
The steps to implement a generalized protocol in a given structure are the same as those given in the preceding section for the  illustrative protocol. The main task in generalization is to find  the requisite sets of multidimensional states  which can implement the task.

\subsection*{Identification of states} 

Consider a network $\kappa$  characterized by the set $\{n,k,\ell_j\}$, where $n$ is the total number of participants, and $k$ is the total number of layers. Let $\ell_j$ be the number of layers to which a participant $u_j$ belongs.  In order to identify the requisite sets of states that can distribute keys simultaneously in this network, we employ the following procedure.
\begin{enumerate}
\item We start with two sets of reference multiqubit states that distribute a key in the $i^{{\rm th}}$ layer,
    \begin{align}\label{eq:layer_i}
    {\cal S}^{(i)}_1:\Big\{\bigotimes_j\ket{0}_{u_j},\bigotimes_j\ket{1}_{u_j}\Big\};~~
    {\cal S}^{(i)}_2:\Big\{\bigotimes_j\ket{+}_{u_j},\bigotimes_j\ket{-}_{u_j}\Big\},
\end{align}
       where $u_j$ corresponds to $j^{\rm th}$ participant in the  $i^{{\rm th}}$ layer.   Similarly,  two sets 
    \begin{align}\label{eq:layer_k}
    {\cal S}^{(k)}_1:\Big\{\bigotimes_l\ket{0}_{u_l},\bigotimes_l\ket{1}_{u_l}\Big\};~~
  {\cal  S}^{(k)}_2:\Big\{\bigotimes_l\ket{+}_{u_l},\bigotimes_l\ket{-}_{u_l}\Big\},
\end{align}
can be employed to distribute a key in the $k^{{\rm th}}$ layer.
     The symbol $u_l$ represents the $l^{\rm th}$ participant belonging to the $k^{{\rm th}}$ layer.
     \item Given the  states in equations (\ref{eq:layer_i}) and (\ref{eq:layer_k}), the two sets that \textit{simultaneously} distribute keys in both the $i^{\rm th}$ and the $k^{\rm th}$ layers with \textit{full confidentiality} are,
        \begin{align}
        {\cal S}_1^{(ik)}: & \Big\{\bigotimes_{jl}\ket{0}_{u_j}\ket{0}_{u_l},\bigotimes_{jl}\ket{0}_{u_j}\ket{1}_{u_l}, \bigotimes_{jl}\ket{1}_{u_j}\ket{0}_{u_l}, \bigotimes_{jl}\ket{1}_{u_j}\ket{1}_{u_l}\Big\}\nonumber\\
        &\equiv {\cal S}_1^{(i)}\otimes {\cal S}_1^{(k)},\nonumber\\
        {\cal S}_2^{(ik)}: & \Big\{\bigotimes_{jl}\ket{+}_{u_j}\ket{+}_{u_l},\bigotimes_{jl}\ket{+}_{u_j}\ket{-}_{u_l}, \bigotimes_{jl}\ket{-}_{u_j}\ket{+}_{u_l}, \bigotimes_{jl}\ket{-}_{u_j}\ket{-}_{u_l}\Big\}\nonumber\\
        &\equiv {\cal S}_2^{(i)}\otimes {\cal S}_2^{(k)}.
         \label{Direct_product_bases}
    \end{align}
    In a similar manner, two sets that \textit{simultaneously} distribute keys in the network $\kappa$ can  be compactly written as:
    \begin{align}\label{eq:kappa}
       {\cal S}_1^{(\kappa)}: & \bigotimes_{r}{\cal S}_1^{(r)},~~~~ {\cal S}_2^{(\kappa)}:~ \bigotimes_{r}{\cal S}_2^{(r)},
    \end{align}
    where $r$ runs over the number of layers. Note that the sets ${\cal S}_1^{(\kappa)}$ and ${\cal S}_2^{(\kappa)}$ contain $\sum_{j}\ell_j$ party qubit states in which the participant $u_j$ has $\ell_j$ qubits.
\item For a given participant $u_j$, we arrange the qubits $\ket{m_1}, \ket{m_2}, \cdots, \ket{m_{l_j}}$ sequentially and employ the bijective mapping,
\begin{equation}\label{eq:mapping}
     \ket{ m_1m_2\cdots m_{l_j}} \leftrightarrow \ket{\sum_{r=1}^{l_j}2^{l_j-r}m_{r}}\equiv \ket{m}_{u_j},\quad m_r\in\{0,1\}.
    \end{equation}
    Thus, states of $\ell_j$ qubits in order with participant $u_j$ is identified to be an equivalent $2^{l_j}$-- dimensional qudit $\ket{m}_{u_j}$.

With this, the two sets in equation (\ref{eq:kappa}) map to two sets $ S_1^{(\kappa )}$ and $ S_2^{(\kappa )}$ having $n$-party multidimensional states that can be employed to share keys \textit{simultaneously} in all the layers of the  network $\kappa$ \textit{confidentially}.
    \end{enumerate}
   
\subsubsection{Key generation rules}
To obtain keys, each participant expresses his outcome in the binary representation. Let $o_j$ be the outcome of the participant $u_j$. Then, in binary representation, it is expressed as:
\begin{equation}
    o_j=\sum_{m=0}^{l_j-1}2^mo_j^{(m)},
\end{equation}
where (recall that) $\ell_j $ is the number of layers to which the participant $u_j$ belongs.
The symbols $o_j^{(m)}$  generate keys in $(\ell_j-m)^{\rm th}$ layer.

\section{Robustness of the protocol}
\label{security}
In this section, we demonstrate the robustness of the protocols against five eavesdropping strategies. For simplicity, it suffices to show them for the illustrative protocol  presented in section (\ref{Illustrative_QKD}) since their generalizations are  straightforward.
In the protocol, Alice sends states to both Bob$_1$ and Bob$_2$. We discuss the two cases one-by-one.

\subsection{Attacks on Bob$_2$}
We first consider the attacks on the states traveling to Bob$_2$. These attacks provide Eve with information shared in layer $L_2$.\\

\noindent{\textbf{(I) Intercept-resend attack}}\\

\noindent In this attack, Eve intercepts the traveling state, performs a measurement, and sends the post-measurement state to the intended recipient. Eve's intervention gets reflected when Alice analyses the data. 
  
Alice sends the states from the two bases $B_1^{(2)}$ and $B_2^{(2)}$ randomly to Bob$_2$. Since both bases are used in key generation, Eve performs random measurements in the two bases on intercepted states.
  The rounds in which her bases match with that of Alice and Bob$_2$, she gains full information without introducing any error. The rounds in which the choice of Eve's basis   does not match with those of Alice, she retrieves the original state only with a probability of $1/2$. In such a case, her presence gets detected with  a probability of $\big(1-\frac{1}{2}\big)$ in one round. 
 So, in $l$ rounds, Eve is detected with a probability of $p_{2}=1-\frac{1}{2^l}$,  which approaches unity for a sufficiently large $l$. \\

\noindent{\textbf{ (II) Entangle-and-measure attack}}\\

\noindent In this attack, Eve employs an entangling operation, which in this case is equivalent to $U\equiv$ CNOT operation (because Alice sends a qubit to Bob$_2$). The effect of $U$ can be expressed as:
 \begin{align}
    & U\ket{0}_T\ket{0}_E=\ket{00}_{TE},~~~ U\ket{1}_T\ket{0}_E=\ket{11}_{TE}.
 \end{align}
 Thus, a disturbance is introduced in those rounds in which Alice sends states from $S_2$ as is also clear from the following equations:
 \begin{align}
     &U\ket{+0}_{TE}=\frac{1}{\sqrt{2}}\big(\ket{++}+\ket{--}\big)_{TE},~~ U\ket{-0}_{TE}=\frac{1}{\sqrt{2}}\big(\ket{+-}+\ket{-+}\big)_{TE}.
 \end{align}
The above equation clearly indicates that Eve introduces errors with a probability of $0.5$ and that too without gaining any information. Thus, the error probability in $l$ such rounds is $(1-\frac{1}{2^l})$.\\

\noindent{\textbf{ (III) Cloning-based attack}}\\

Suppose that Eve interacts with states traveling to Bob$_2$ with a unitary operation $U_E$, whose action is as follows,
\begin{align}\label{eq:cloning_qubit}
&\ket{i}_T\ket{0}_E\xrightarrow{U_E} \sqrt{F}\ket{iE_{ii}}_{TE}+\sqrt{D}\ket{jE_{ij}}_{TE},~~~~~i\neq j,~i,j\in\{0,1\},
\end{align}
where $F=1-D$ is the probability with which Bob$_2$ gets the correct result.
The subscripts $T$ and $E$ represent the  traveling state and Eve's state respectively. In such a case, Eve introduces an error with a probability of $D$. Thus, Eve will be detected with a probability $D^l$ for $l$ rounds. 
For the cloning-based attack given in equation (\ref{eq:cloning_qubit}), 
mutual information between Alice and Bob is given by,
\begin{align}
    I_{AB_2} =1-h(F),
\end{align}
where $h(x)=-x\log_2{x}-(1-x)\log_2(1-x)$ is the binary entropy. Since Eve wants to clone both bases equally, she can do that with an optimal fidelity \cite{cerf2002security} $ F_E =\frac{1}{2}+\sqrt{1-F}$. Thus, the mutual information between Alice and Eve is,
\begin{align}
    I_{AE} = 1-h(F_E).
\end{align}
 Eve gains information at the cost of introducing errors in the outcomes of Alice and Bob$_2$. Thus, the proposed protocols are robust against such eavesdropping.

\subsection{Attack on Bob$_1$}
Since Bob$_1$ belongs to both layers $L_1$ and $L_2$, depending upon Eve's attacks, she may retrieve information being shared either in one layer or in both layers.\\
 
\noindent{\textbf{Eavesdroppiong in a  single layer}}\\

We first dispose of a very simple attack in which Eve attempts to know the key in any single layer, say, $L_2$. In that case, she would employ a  degenerate observable
 \begin{align}
     O\equiv \lambda_1\big(\ket{0}\bra{0}+\ket{2}\bra{2}\big)+\lambda_2\big(\ket{1}\bra{1}+\ket{3}\bra{3}\big),~~~~~~\lambda_1\neq\lambda_2,
 \end{align}
on the states traveling from Alice to Bob$_1$.  The effect of this attack is equivalent to an attack in QKD protocols running for a single layer. For these attacks, the security analyses done in \cite{shor2000simple,cerf2002security,mafu2022security} may be imported to this protocol.\\

\noindent{\textbf{Eavesdropping in both the layers}}\\

 The more vicious attacks are those in which Eve exploits the layered nature of the network which has no counterpart elsewhere. The hierarchy in the layers allows Eve to identify the more preferred layer (layer $L_1$ in this case). Eve's attacks on this layer provide her with information being shared in all the layers to which these participants belong (layers $L_1$ and $L_2$ in this case). \\
 
 \noindent{\textbf{ (I) Intercept-resend attack}}\\

 In this attack, as the name suggests, Eve intercepts the traveling state, performs a measurement, and sends the post-measurement state to the intended recipient. Eve's intervention gets reflected when Alice analyses the data. The following analysis shows how the hierarchy in layers provides Eve with the information being shared in other layers as well. 
  
In the protocol, Alice sends the states from the two bases $B_1^{(1)}$ and $B_2^{(1)}$ randomly to Bob$_1$. Since both bases are used in key generation, Eve performs random measurements in the two bases on intercepted states.
  The rounds in which her bases match with those of Alice and Bob$_1$, she gains complete information without introducing any errors. The rounds in which Eve's choice of basis does not match with that of Alice, she retrieves the original state only with a probability of $1/4$. In such a case, her presence gets detected with  a probability of $\big(1-\frac{1}{4}\big)$ in one round.
   Thus, in $l$ rounds, the presence of Eve can be detected with a probability of $p_{1}=1-\frac{1}{4^l}$ which approaches unity for a sufficiently large\footnote{If a participant belongs to $l_j$ layers, then  the probability of detecting Eve is given by $1-d_j^{-l}$;~ $d_j=2^{l_j}$.} $l$.  Let $1-\epsilon$ be the desired probability of eve's detection, then,
 \begin{align}
     1-\frac{1}{4^l} = 1-\epsilon\implies l\approx o\Big(\log\frac{1}{\epsilon}\Big).
 \end{align}
 Clearly, the number of rounds grows logarithmically with $\epsilon^{-1}$.
  
  In this attack, whenever Eve's measurement basis matches with that of each Bob, she obtains full information about the key. Since Bob$_1$ measures a ququart, Eve  gets the same result as Bob$_1$ and hence obtains information being shared in both layers. \\ 

\noindent{\textbf{ (II) Entangle-and-measure attack}}\\

 In this attack, Eve would entangle her ancillary systems with traveling subsystems and then try to retrieve information by performing measurements on them.

 Consider an entangle-and-measure attack in the first layer. In this case, Eve would entangle her ancillary system with the subsystem traveling to Bob$_1$. The effect of this operation can be expressed as:
 \begin{align}
    & U_1\ket{i}_T\ket{0}_E\equiv\ket{ii}_{TE}.
 \end{align}
 The subscripts $T$ and $E$ represent the traveling subsystem and Eve's subsystem respectively. 
 For the rounds in which Alice has sent states from the set $S_1$, no disturbance will be produced and Eve obtains full information. However, Alice also sends states from the set $S_2$ randomly. The effect of Eve's  intervention on these states can be expressed as:
 \begin{align}
     &U_1\ket{0'}_T\ket{0}_E=\frac{1}{2}\big(\ket{0'0'}+\ket{1'1'}+\ket{2'2'}+\ket{3'3'}\big),\nonumber\\
      &U_1\ket{1'}_T\ket{0}_E=\frac{1}{2}\big(\ket{0'1'}+\ket{1'0'}+\ket{2'3'}+\ket{3'2'}\big),\nonumber\\
       &U_1\ket{2'}_T\ket{0}_E=\frac{1}{2}\big(\ket{0'2'}+\ket{1'3'}+\ket{2'0'}+\ket{3'1'}\big),\nonumber\\
        &U_1\ket{3'}_T\ket{0}_E=\frac{1}{2}\big(\ket{0'3'}+\ket{1'2'}+\ket{2'1'}+\ket{3'0'}\big).
 \end{align}
The above equation reflects the fact that Eve introduces an error with a probability of $\frac{3}{4}$, and does not obtain any information. Thus, in $l $ such rounds, the probability of detection of Eve is $\big(1-\frac{1}{4^l}\big)$. \\
 
 \noindent{\textbf{(III) Cloning-based attack}}\\
 \label{Secret key rate analysis}

 In this attack, Eve couples her ancilla with traveling states with a special interaction that copies two mutually unbiased bases equally. Let $U_E$ be the unitary operation that realizes this task.\\ 

In layer $L_1$, Bob$_1$ is the only participant to whom states are transmitted. The  action of $U_E$ on the ququart transmitted to Bob$_1$ can be expressed as:
\begin{align}\label{eq:Eve}
&\ket{i}_T\ket{0}_E\xrightarrow{U_E} \sqrt{F}\ket{iE_{ii}}_{TE}+\sqrt{\frac{D}{3}}\Big(\sum_{j\neq i}\ket{jE_{ij}}_{TE}\Big),~~~~~~~i,j\in\{0,1,2,3\},
\end{align}
where $F=1-D$ is the probability with which Bob$_1$ gets the correct result. The states $\ket{E_{ij}}$ represent the states of Eve's ancilla. So, the mutual information between Alice and Bob$_1$ is:
\begin{align}\label{eq:IAB1}
    I_{AB_1} =2+F\log_2(F)+(1-F)\log_2\Big(\frac{1-F}{3}\Big).
\end{align}
Following equation (\ref{eq:Eve}), Eve's interventions introduce errors with a probability $D$  which are reflected when Alice and Bob$_1$ compare subsets of their outcomes. As a result of these interventions, the mutual information between Alice and Bob$_1$ also decreases as is reflected in equation (\ref{eq:IAB1}).

Since Eve wishes to obtain information from two bases equally, she can do this with optimal fidelity \cite{cerf2002security}, $F_E=\frac{3}{4}-\frac{F}{2}+\frac{\sqrt{3(1-F)}}{2}$. In that case, the mutual information between Alice and Eve is,
\begin{align}\label{eq:IAE}
    I_{AE} = 2+F_E\log_2(F_E)+(1-F_E)\log_2\Big(\frac{1-F_E}{3}\Big).
\end{align}
Thus, $I_{AE}$ gives information gained by Eve when she introduces errors with a probability $D$. If Eve does not introduce any error, her information gain is zero, thereby reflecting the resilience of protocols against such attacks.

The information shared between Alice and Bob$_1$ is much larger than the information shared between Alice and Bob$_2$. This is because Bob$_1$ belongs to two layers in contrast to Bob$_2$. This hierarchy in layers is also reflected in the information gain of Eve which is much larger for layer $L_1$ in contrast to when Eve attacks Bob$_2$'s subsystem.\\
\subsection{Photon-number-splitting attack}\label{photon}
 Prepare-and-measure protocols are mostly implemented with weak coherent pulses (WCP), which provide Eve with additional opportunities. In such cases, Eve  may block all the single photons by employing photon counters, and split the multi photons into two parts\footnote{We should keep in mind that WCP is not an eigenstate of the number operator.}. She sends one of them to the intended recipient and keeps the other ones with herself. Afterward, by using the data shared in  public communication, Eve measures the states with herself in the same basis as used by participants, thus retrieving full information without introducing any errors. Such kinds of attacks can be circumvented by employing additional WCP with lower mean photon numbers as done in  \cite{wang2005beating}. Due to the random transmission of two WCPs, Eve cannot distinguish them and she  blocks all single-photon pulses. Due to this, Bob will get less number of pulses for additionally used WCPs that reflect the presence of an eavesdropper.

\subsection{Trojan horse attack}\label{Trojan}
 The current proposals of protocols are vulnerable to Trojan horse attacks such as `the delay photon attack' and `the invisible photon attack'.  However, with slight modifications  \cite{qin2019quantum}, our protocol can be made robust against these attacks. To prevent the delayed photon attack, Bob should randomly choose a subset of signals to count for the number of photons using a photon number splitter. This data compared with the one that is employed to send signals reveals the presence of such attacks if any. Invisible photon attacks can be countered by using a filter that allows  photons of only a specified wavelength  to reach the system. \\
Different eavesdropping attacks and their remedies have been compactly shown in table (\ref{tab:my_label}).
\begin{table}[h!]
    \centering
    \resizebox{!}{1.1cm}{
\begin{tabular}{||c| c ||} 
 \hline
Attack & Remedy  \\ [0.5ex] 
 \hline\hline
 Intercept-resend attack & Nonorthogonal states \\ 
 \hline
 Entangle-and-measure attack & Nonorthogonal states \\
 \hline
  Cloning based attack & Nonorthogonal states \\
 \hline
 Photon-number-splitting attack & Decoy pulses \\
 \hline
 Trojan horse attack &  Insertion of photon number splitter\\
 \hline
 \hline
\end{tabular}}
    \caption{Different attacks on the layered QKD protocol and the corresponding remedies.}
    \label{tab:my_label}
\end{table}

\subsection{Information-theoretic security}
\label{Information-theoretic_security_LQKD}
In the above analysis, we have shown the robustness of the protocol against several eavesdropping strategies. Information-theoretic security analysis of the proposed protocol follows directly from the information-theoretic security analyses for the individual layers. This owes to the mathematical equivalence between (I)  the proposed protocol (with multidimensional separable states) and (II)  two QKD protocols running in parallel (with two-dimensional separable states). This feature is also reflected in the procedure for the identification of resource states (laid down in section (\ref{generalisation})). For example, consider the resource states given in equation (\ref{eq:sets}) for the protocol in section (\ref{Illustrative_QKD}), which we rewrite here for the sake of completeness,
\begin{align*}
    & S_1: \Big\{\ket{00}_{b_1b_2}, \ket{11}_{b_1b_2}, \ket{20}_{b_1b_2}, \ket{31}_{b_1b_2}\Big\},\nonumber\\
    & S_2: \Big\{\ket{0'+}_{b_1b_2}, \ket{1'-}_{b_1b_2}, \ket{2'+}_{b_1b_2}, \ket{3'-}_{b_1b_2}\Big\}.
   \end{align*}
   In the above equations, the subscripts $b_1$ and $b_2$ refer to the states corresponding to Bob$_1$ and Bob$_2$  respectively.
   Employing the decimal-to-binary mapping given in equation (\ref{eq:mapping}), 
   \begin{equation*}
     \ket{ m_1m_2\cdots m_{l_j}} \leftrightarrow \ket{\sum_{r=1}^{l_j}2^{l_j-r}m_{r}}\equiv \ket{m}_{u_j},\quad m_r\in\{0,1\},
    \end{equation*}
     for the four-level states, the two sets can be written as:
\begin{align*}
   & {\cal S}_1: \Big\{\ket{00}_{b_1}\ket{0}_{b_2}, \ket{01}_{b_1}\ket{1}_{b_2}, \ket{10}_{b_1}\ket{0}_{b_2}, \ket{11}_{b_1}\ket{1}_{b_2}\Big\},\nonumber\\
    &{\cal S}_2: \Big\{\ket{++}_{b_1}\ket{+}_{b_2}, \ket{+-}_{b_1}\ket{-}_{b_2}, \ket{-+}_{b_1}\ket{+}_{b_2}, \ket{--}_{b_1}\ket{-}_{b_2}\Big\},
   \end{align*}
It is clear from the above equations that Bob$_1$ is in the possession of two qubits. The second qubit of Bob$_1$ and  the qubit of Bob$_2$, chosen uniformly from  the sets $\{\ket{00}_{b_1b_2}, \ket{11}_{b_1b_2}\}$ and $\{\ket{++}_{b_1b_2}, \ket{--}_{b_1b_2}\}$,  distribute a key in layer $L_2$. The first qubit of Bob$_1$, chosen uniformly from the sets $\{\ket{0}_{b_1}, \ket{1}_{b_1}\}$ and $\{\ket{+}_{b_1}, \ket{-}_{b_1}\}$, distributes a key in layer $L_1$. This is nothing but two QKD protocols running in parallel.

Therefore, the information-theoretic security analysis of the proposed protocol may be performed by combining the information-theoretic security analyses of QKD between the two participants (constituting the layer $L_1$)\cite{shor2000simple} and among the three participants \cite{matsumoto2007multiparty} (constituting the layer $L_2$). Extending this prescription for $k$ layers will ensure the security of key distribution in general networks. 
\section{Pinpointing Eve}\label{pinpoint}
The preceding section shows the robustness of protocols presented in section (\ref{Illustrative_QKD}) which distributes secure keys in all layers of a network simultaneously. Confidentiality of keys in distinct layers allows sharing of secure keys in some of the layers, even if there is eavesdropping in other layers. However, this would require pinpointing the location of Eve. For this purpose, Alice analyses errors introduced in the outcomes of all the Bobs separately. This analysis helps in the identification of Bob whose data is being intercepted by Eve and hence all the layers in which he is a participant can be excluded. Keys shared in the rest of the layers remain secure. To illustrate it, consider an example of the protocol given in section (\ref{Illustrative_QKD}). Suppose that Eve intercepts states traveling to Bob$_2$. In such a case, the correlation between Alice and  Bob$_2$ decreases whereas that between Alice and Bob$_1$ remains intact. This shows that information shared in layer $L_1$ (which does not have Bob$_2$ as a participant) is secure in spite of the fact that the key in layer $L_2$ is to be discarded.

This concludes our discussion of layered QKD protocols. We now move on to layered SQKD protocols.
\section{Prepare-and-measure SQKD protocol in  layered networks}
\label{PandMlayeredSQKD} 
The idea of SQKD, as already discussed in the introduction, allows for the secure distribution of keys with minimal quantum resources. This particular feature not only answers the fundamental question of  minimum quantum resources needed for QKD  but also eases down the experimental implementation of the protocol \cite{massa2022experimental}. The SQKD protocol proposed in \cite{Boyer07} involves only two participants. However, the realistic scenarios correspond to layered networks having multiple participants distributed in different layers. Keeping this in mind, in this section, we propose an SQKD protocol that allows  the distribution of keys simultaneously in all the layers of the network with only one quantum participant. However, before going to that, we make a slight detour and briefly recapitulate the SQKD protocol proposed in \cite{Boyer07} to share a secure key between two participants for the purpose of pedagogy in the next section.
\subsection{Brief recapitulation of SQKD protocol \cite{Boyer07} between two participants}

 In \cite{Boyer07}, an SQKD protocol has been proposed that securely distributes a  key between a quantum (Alice) and a classical (Bob) participant. Recall that a quantum participant can prepare any state and perform measurements in any basis whereas a classical participant can prepare and measure only in the computational basis. The steps of the protocol are indicated in the figure (\ref{LSQKD}) and explicitly listed in Appendix (\ref{SQKD_Boyer}).
\begin{figure}[h]
    \centering    \includegraphics[width=10cm
    ]{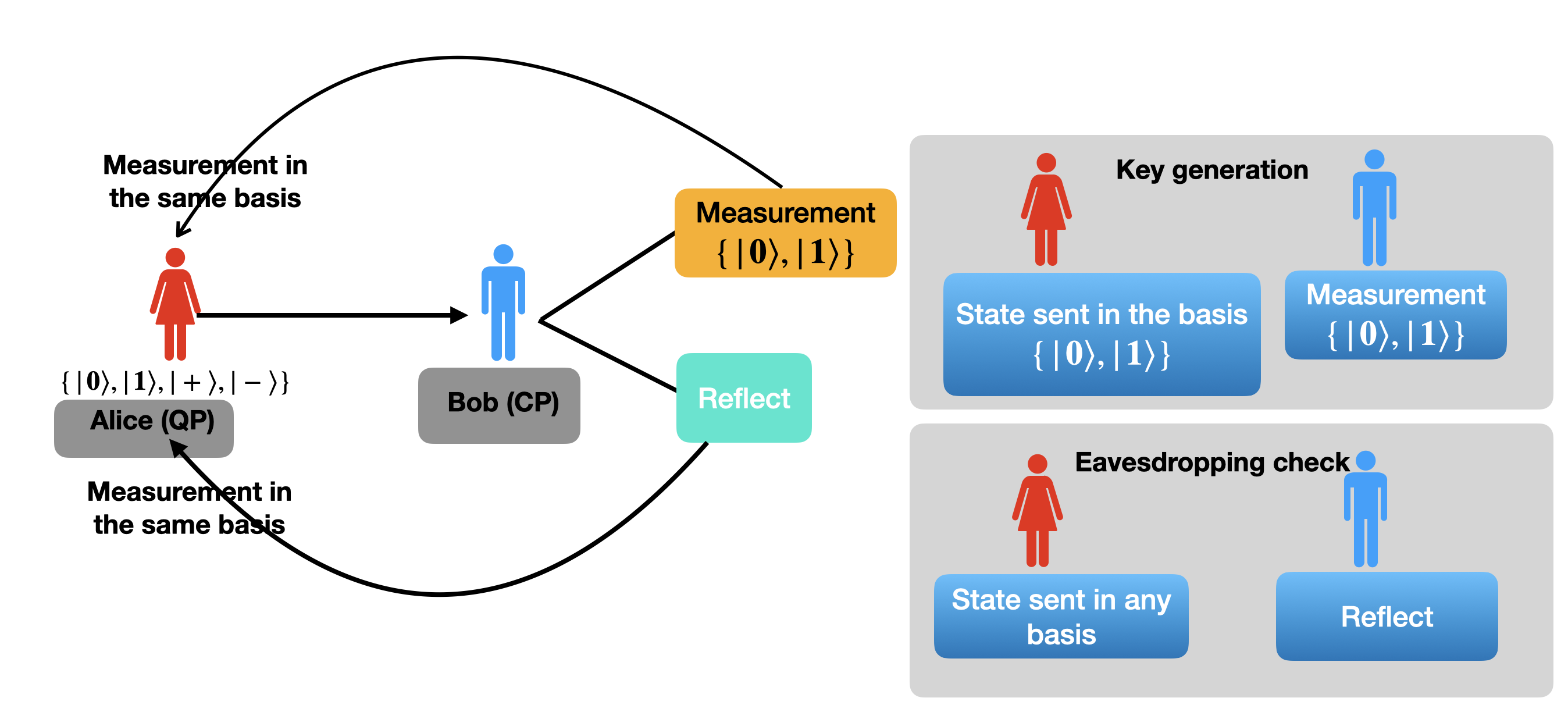}
    \caption{Pictorial representation of the prepare-and-measure SQKD protocol. Alice is the quantum participant (QP) and Bob is the classical participant (CP). }
    \label{LSQKD}
\end{figure}

We now move on to propose prepare-and-measure SQKD protocols that simultaneously  distribute keys in a layered network.

\subsection{Prepare-and-measure SQKD protocol in layered networks}
\label{SQKD}
We adopt the same strategy as has been adopted for QKD.  In this section, we provide a key distribution protocol that can distribute keys simultaneously in a layered network with only one quantum participant\footnote{An example of such a protocol for a different network may be found in our recent work \cite{bala2022semi}  in its preliminary form without security analysis. However, in this work, we present the protocol together with security analysis and procedure for generalization to arbitrarily layered networks.}. For simplicity, we consider the same network as has been considered for layered QKD in section (\ref{Illustrative_QKD}).\\ 
\noindent\textbf{\textit{Aim:} } Simultaneous sharing of keys in layers $L_1$ (Alice, Bob$_1$) and $L_2$ (Alice, Bob$_1$, Bob$_2$) with Alice being the only quantum participant.\\
\noindent\textbf{\textit{Resources:}} The two sets of $4\otimes 2$ dimensional separable states,  
\begin{align*}
    S_1: \Big\{\ket{00}, \ket{11}, \ket{20}, \ket{31}\Big\},~ ~S_2: \Big\{\ket{0'+}, \ket{1'-}, \ket{2'+}, \ket{3'-}\Big\},~
   \end{align*}
   also serve the purpose here, thanks to the same structure of the network. The states in the set $S_2$ are Fourier transforms of states in the set $S_1$ and are defined as:
     \begin{align*}
       \ket{j'} \equiv \frac{1}{2}\sum_{k=0}^3e^{2\pi ijk/4}\ket{k},~j'\in\{0,1,2,3\}, ~{\rm and}~\ket{\pm} = \frac{1}{\sqrt{2}}\big(\ket{0}\pm \ket{1}\big).
   \end{align*}
\textit{Measurement basis of Bob$_1$:}   $B_1^{(1)} \equiv \{\ket{0}, \ket{1}, \ket{2}, \ket{3}\}$,\\
\textit{Measurement basis of Bob$_2$:} $B_1^{(2)} \equiv \{\ket{0}, \ket{1}\}$.
\begin{figure}[h]
    \centering
    \includegraphics[width=11cm]{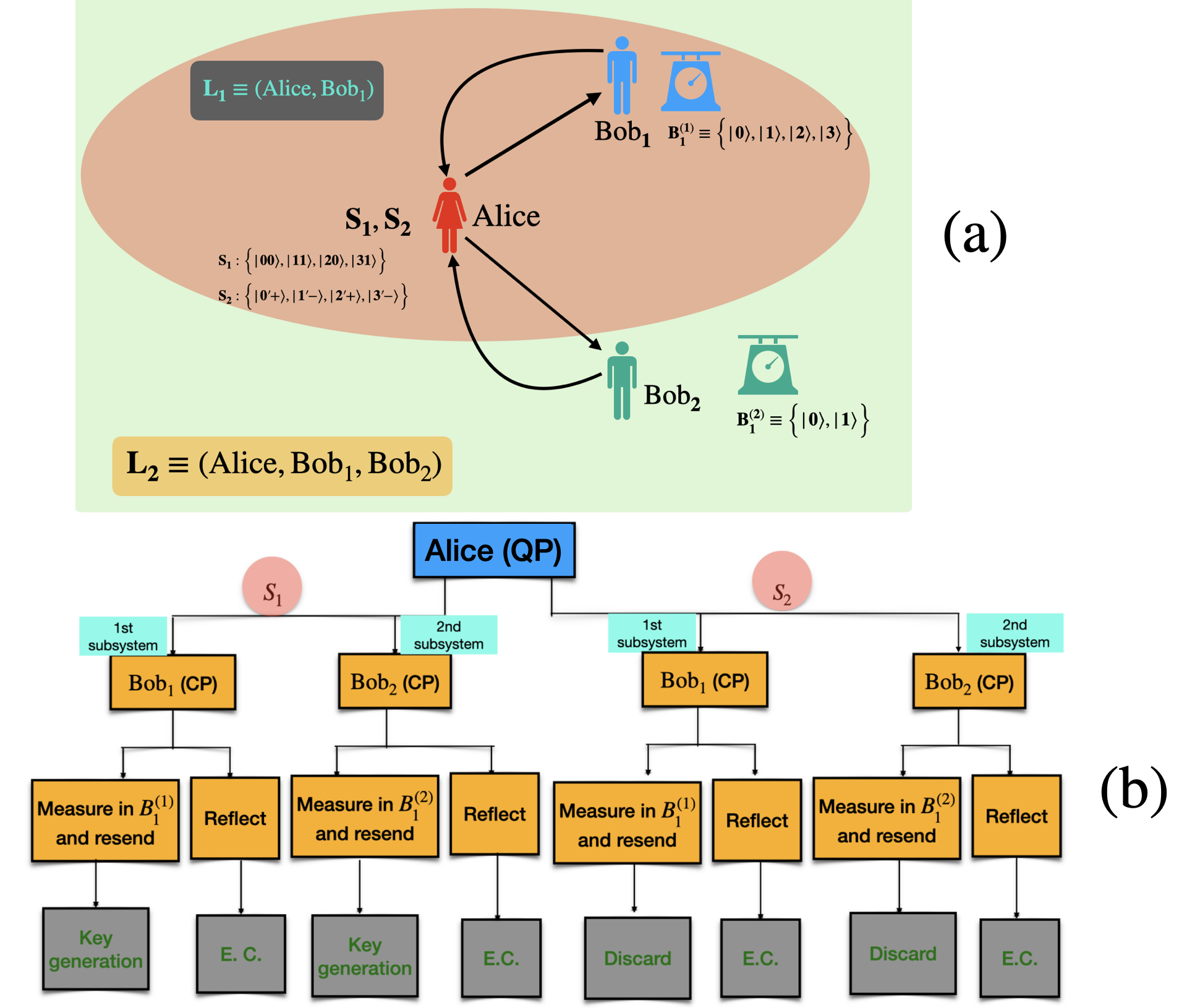}
    \caption{Schematic diagram of the  layered SQKD protocol. Figure (a) provides a description of the network and requisite resources for the task. Figure (b) illustrates the steps of the protocol. QP, CP, and E.C. represent quantum participant, classical participant, and eavesdropping check respectively.}
    \label{fig:LSQKD}
\end{figure}
 \subsection*{The protocol}
 The steps of the protocol are as follows:
\begin{enumerate}
    \item Alice randomly prepares a state with an equal probability from  ${S}_1$ or ${S}_2$ and sends the first and the second subsystems to Bob$_1$ and Bob$_2$ respectively. She also records the quantum number of the first subsystem in base $2$ representation.
    \item Each Bob, being a classical participant, exercises two choices with equal probabilities. He either measures the received state in the computational basis  and sends the post-measurement state to Alice, or simply returns the received state to Alice. 
    \item On receiving the states from each Bob, Alice measures them in the same bases in which they were initially prepared. 
    \item This process is repeated for at least $8n(1+\delta)$ rounds, where $n$ is the length of the key to be shared and $\delta>0$ depends on the length of the subset that is used for post-processing.
    After this, on an authenticated classical channel, Alice reveals the rounds in which she has sent states from set $S_1$ and both the Bobs reveal the rounds in which they have measured.
    \item {\it Eavesdropping check:} Employing this information, Alice analyses data of the rounds in which either Bob$_1$ or Bob$_2$ or both have not performed any measurement. In the absence of eavesdropping, Alice would get the same state as she had sent. 
    \item \textit{Key generation:} If eavesdropping is so ruled out, the data of those rounds in which Alice has sent a state from $S_1$ and both the Bobs have performed measurements constitute a key in both the layers. In addition, the rounds in which Bob$_1$ has measured (irrespective of the measurement of Bob$_2$) also constitute key symbols in layer $L_1$. 
    \end{enumerate}
    The structure of the network, resources needed, and the steps of the protocol are compactly shown in the figure (\ref{fig:LSQKD}).
    
    \subsubsection*{\textbf{Key generation rule}}
To retrieve a key symbol in each layer, the same strategy  as employed in layered QKD (please see table (\ref{tab:key_rule_QKD})) is used with the sole difference that only the  set $S_1$ contributes to key generation.

The generalization of this protocol to arbitrary networks is straightforward. The steps of the protocol essentially remain the same. The main task is  the identification of appropriate resource states, the procedure for which is provided in the section (\ref{generalisation}).

 We now move on to discuss the robustness of the proposed layered SQKD protocol against various eavesdropping attacks.

 \section{Robustness of the protocol}
 \label{security1}
 Layered SQKD protocol is different from  layered QKD protocol in that it has only one quantum participant and requires two-way communication. The robustness of the former against one-way eavesdropping attacks follows directly from that of the latter. In the following, we show the robustness of layered SQKD protocol against a two-way eavesdropping attack which has been considered for other semi-quantum protocols as well \cite{li2013quantum}.
 
 \subsection{Security against two-way entangling attacks}
 In these attacks, Eve entangles her ancillae, initially in states $\ket{0}_F$ and $\ket{0}_B$ with the traveling states in both the forward and backward paths with unitaries $U_F$ and $U_B$ respectively.  
\subsubsection{Attack on Bob$_2$'s subsystem}
Since Alice sends qubits to Bob$_2$, actions of Eve's interactions on these states can be described as:
 \begin{align}
 &U_F\ket{i}_{T}\ket{0}_{F}\equiv \sum_{j=0}^1\ket{j}_T\ket{E_{ij}}_F,~~U_B\ket{i}_{T}\ket{0}_{B}\equiv \sum_{j=0}^1\ket{j}_T\ket{F_{ij}}_B,\label{eq:qubit_B}
 \end{align}
  where the respective states $\ket{E_{ij}}$ and $\ket{F_{ij}}$ are unnormalized states of  the first and the second ancillae  of Eve satisfying the constraints imposed by unitary operations.
Since Bob$_2$ has two choices- we consider the two cases separately as follows: 
 {\subsubsection*{Case I:  Bob$_2$ chooses measure-and-resend}} Consider a particular round in which Alice has sent a state $\ket{0}$ to Bob$_2$ and Bob$_2$ performs a measure-and-resend operation on the state.
 In that case, following the action of $U_F$ given in equation (\ref{eq:qubit_B}), the combined state of the traveling system and Eve, after Bob$_2$'s measurement will be,
 \begin{align}
     &\ket{0}\ket{E_{00}},~~~~~~{\rm OR}~~\ket{1}\ket{E_{01}}.
 \end{align}
  Eve interacts with the traveling state with her ancilla through unitary interaction $U_B$. 
  Then, the combined states of the traveling system and both of Eve's ancillae can be expressed as:
  \begin{align}  
  &U_B\ket{0}_T\ket{E_{00}}_F\ket{0}_{B}=\sum_{j=0,1}\ket{j}_T\ket{F_{0j}}_B\ket{E_{00}}_F,\\
  &~~~~~~~~~~~~~~~~~~~~~~~~~~~~~~{\rm OR}\nonumber\\
  &U_B\ket{1}_T\ket{E_{01}}_F\ket{0}_{B}=\sum_{j=0,1}\ket{j}_T\ket{F_{1j}}_B\ket{E_{01}}_F.
  \end{align}
  Alice, upon receiving the states, perform a measurement in the computational basis. She receives the same post-measurement state as she had sent with a probability  $p=\norm{\ket{E_{00}}}^2\cdot\norm{\ket{F_{00}}}^2+\norm{\ket{E_{01}}}^2\cdot\norm{\ket{F_{10}}}^2$. The first term corresponds to the complete correlation between  the two parties, however, the second term corresponds to anti-correlation between the two parties leading to the detection of eavesdropper when a subset of data is made public. The other scenario in which Alice gets a post-measurement state different from what she had sent also reveals the presence of an eavesdropper with a  probability equal to $(1-p)$.

 A similar analysis holds when Alice sends the state $\ket{1}$ to Bob$_2$. Probabilities 
of Eve's detection have been given explicitly  in the table (\ref{Two_SQKD}).
\begin{table}[h!]
 \begin{center}
 \resizebox{!}{2.25cm} {
    \centering
\begin{tabular}{|c | c| c| c |c|}
\hline
\multirow{3}{2cm}{State sent by Alice to Bob$_2$}&\multirow{3}{2cm}{\centering Bob$_2$'s post-measurement state}&\multirow{3}{1.75cm}{Alice's post-measurement state} &\multirow{3}{1.85cm}{Correlation}&\multirow{3}{1.85cm}{detcetion probability of Eve}\\
 &&&& \\
&&&&\\\hline
\multirow{2}{*}{$\ket{0}$} &\multirow{2}{*}{$\ket{0}$}&\multirow{2}{*}{$\ket{0}$}&\multirow{2}{*}{$\checkmark$}&\multirow{2}{*}{$0$}\\
&&&&\\\hline
\multirow{2}{*}{$\ket{0}$} &\multirow{2}{*}{$\ket{0}$}&\multirow{2}{*}{$\ket{1}$}&\multirow{2}{*}{$\boldsymbol{\times}$}&\multirow{2}{*}{$\norm{\ket{E_{00}}}^2\cdot\norm{\ket{F_{01}}}^2$}\\
&&&&\\\hline
 \multirow{2}{*}{$\ket{0}$}&\multirow{2}{*}{$\ket{1}$}& \multirow{2}{*}{$\ket{0}$}&\multirow{2}{*}{$\boldsymbol{\times}$}&\multirow{2}{*}{$\norm{\ket{E_{01}}}^2\cdot\norm{\ket{F_{10}}}^2$}\\
&&&&\\\hline
\multirow{2}{*}{$\ket{0}$}&\multirow{2}{*}{$\ket{1}$}&\multirow{2}{*}{$\ket{1}$}&\multirow{2}{*}{$\boldsymbol{\times}$}&\multirow{2}{*}{$\norm{\ket{E_{01}}}^2\cdot\norm{\ket{F_{11}}}^2$}\\
&&&&\\ \hline
\multirow{2}{*}{$\ket{1}$}&\multirow{2}{*}{$\ket{0}$}&\multirow{2}{*}{$\ket{0}$}&\multirow{2}{*}{$\boldsymbol{\times}$}&\multirow{2}{*}{$\norm{\ket{E_{10}}}^2\cdot\norm{\ket{F_{00}}}^2$}\\
&&&&\\ \hline
\multirow{2}{*}{$\ket{1}$}&\multirow{2}{*}{$\ket{0}$}&\multirow{2}{*}{$\ket{1}$}&\multirow{2}{*}{$\boldsymbol{\times}$}&\multirow{2}{*}{$\norm{\ket{E_{10}}}^2\cdot\norm{\ket{F_{01}}}^2$}\\
&&&&\\ \hline
\multirow{2}{*}{$\ket{1}$}&\multirow{2}{*}{$\ket{1}$}&\multirow{2}{*}{$\ket{0}$}&\multirow{2}{*}{$\boldsymbol{\times}$}&\multirow{2}{*}{$\norm{\ket{E_{11}}}^2\cdot\norm{\ket{F_{10}}}^2$}\\
&&&&\\ \hline
\multirow{2}{*}{$\ket{1}$}&\multirow{2}{*}{$\ket{1}$}&\multirow{2}{*}{$\ket{1}$}&\multirow{2}{*}{$\checkmark$}&\multirow{2}{*}{$0$}\\
&&&&\\ \hline
 \end{tabular}
 \vspace{0.3cm}
    \label{tab:general}
    }
    \end{center}
    \caption{Effects of Eve's intervention and probability of Eve's detection. }
    \label{Two_SQKD}
\end{table}

\subsection*{Case II: Bob$_2$ chooses Reflect} The data of these rounds are employed to check for the presence of an eavesdropper. Since Bob$_2$ does not perform any measurement, in the ideal scenarios, the state should have remained  undisturbed and Alice's statistics should have matched before and after the transmission of the state. However, due to Eve's interventions, the statistics change,  resulting in the detection of an eavesdropper. The following two cases arise:
\subsection*{(i) Alice sends a  state from the computational basis} Consider a round in which Alice sends a state $\ket{0}$ or $\ket{1}$ to Bob$_2$. The combined state of the traveling system and that of Eve after the action of two unitaries can be expressed as:
\begin{align}   
\ket{i}_T\ket{00}_{FB}\xrightarrow{U_BU_F}
&\ket{i}_T\big(\ket{E_{ii}}\ket{F_{ii}}+\ket{E_{i\Bar{i}}}\ket{F_{\Bar{i}i}}\big)_{FB}\nonumber\\
&+\ket{\Bar{i}}_T\big(\ket{E_{ii}}\ket{F_{i\Bar{i}}}+\ket{E_{i\Bar{i}}}\ket{F_{\Bar{i}\Bar{i}}}\big)_{FB}~,
\end{align}
${\rm where}~i\in\{0,1\},~\Bar{i}=(i+1)~{\rm mod}~2$. Whenever Alice gets a different state from what she had sent, the presence of Eve is reflected.
Thus, Eve's interventions are detected with a probability of $\norm{\ket{E_{00}}\ket{F_{01}}+\ket{E_{01}}\ket{F_{11}}}^2$ or $\norm{\ket{E_{10}}\ket{F_{00}}+\ket{E_{11}}\ket{F_{10}}}^2$ when Alice sends the states  $\ket{0}$ or $\ket{1}$ respectively. 
\subsection*{ (ii) Alice sends a state from the conjugate basis} Alice may send either of the states, $\ket{+}$ or $\ket{-}$. In these cases, the combined state of the traveling system and that of Eve can be expressed as:
\begin{align}    
\ket{\pm}_T\ket{0}_{F}\ket{0}_{B}\xrightarrow{U_BU_F}
&\ket{+}_T\bigg(\ket{E'_{\pm 0}}\ket{F'_{0+}}+\ket{E'_{\pm 1}}\ket{F'_{1+}}\bigg)_{FB}\nonumber\\
&+\ket{-}_T\bigg(\ket{E'_{\pm 0}}\ket{F'_{0-}}+\ket{E'_{\pm 1}}\ket{F'_{1-}}\bigg)_{FB},
\end{align}
where the states $\ket{E'_{\pm 0}}$, $\ket{E'_{\pm 1}}$, $\ket{F'_{0\pm }}$ and $\ket{F'_{1\pm }}$ are defined as:
\begin{align}
    &\ket{E'_{\pm 0}}\equiv\frac{1}{\sqrt{2}}\big(\ket{E_{00}}\pm \ket{E_{10}}\big),~~~~~\ket{E'_{\pm 1}}\equiv\frac{1}{\sqrt{2}}\big(\ket{E_{01}}\pm \ket{E_{11}}\big),\nonumber\\
    &\ket{F'_{0\pm }}\equiv\frac{1}{\sqrt{2}}\big(\ket{F_{00}}\pm\ket{F_{01}}\big),~~~~~~\ket{F'_{1\pm }}\equiv\frac{1}{\sqrt{2}}\big(\ket{F_{10}}\pm\ket{F_{11}}\big).
\end{align}
Measurement of Alice reveals the presence of an eavesdropper with a probability of $\norm{\ket{E'_{+0}}\ket{F'_{0-}}+\ket{E'_{+1}}\ket{F'_{1-}}}^2$ and $\norm{\ket{E'_{-0}}\ket{F'_{0+}}+\ket{E'_{-1}}\ket{F'_{1+}}}^2$ when the states $\ket{+}$ and $\ket{-} $ are sent to Bob$_2$  respectively.
 \subsubsection{Attack on Bob$_1$'s subsystem}
 Since Bob$_1$ belongs to both the layers, Alice sends ququarts to Bob$_1$. Suppose that actions of Eve's interactions $U_F$ and $U_B$ on travelling ququarts and  her two ancillae $\ket{0}_F$ and $\ket{0}_B$  can be expressed as:
 \begin{align}
          &U_F\ket{i}_{T}\ket{0}_F\equiv \sum_{j=0}^3\ket{j}_T\ket{E_{ij}}_F, ~~~~U_B\ket{i}_{T}\ket{0}_B\equiv \sum_{j=0}^3\ket{j}_T\ket{F_{ij}}_B,
 \end{align}
 where $\ket{E_{ij}}_F$ and $\ket{F_{ij}}_B$ are un-normalized states of Eve's first and second ancilla satisfying the constraints imposed by unitary operations. The following cases may arise:
 \subsection*{Case I: Bob$_1$ chooses measure-and-resend} The data of these rounds are used for key generation, however, a subset of data can be chosen to check for the presence of an eavesdropper. \\
 Consider a round in which Alice sends the state $\ket{0}$ to Bob$_1$. 
 In this case, after the action of $U_F$, the  combined state of the traveling ququart and Eve, after Bob$_1$'s measurement, can be any one of the following,
 \begin{align}
     &\ket{0}_T\ket{E_{00}}_F,~~~{\rm OR}~~\ket{1}_T\ket{E_{01}}_F,~~{\rm OR}~~\ket{2}_T\ket{E_{02}}_F,~~{\rm OR}~~\ket{3}_T\ket{E_{03}}_F.
 \end{align}
 Eve interacts with her other ancilla  the  ququart traveling back to Alice through the unitary $U_B$. The combined state of traveling ququart and the two ancillae of Eve, after this unitary, can be expressed as:
  \begin{align}  &U_B\ket{i}_T\ket{E_{0i}}_{F}\ket{0}_{B}=\sum_{j=0}^3\ket{j}_T\ket{E_{0i}}_{F}\ket{F_{ij}}_{B},~~~~~~~i\in\{0,1,2,3\}.
  \end{align}
  Alice, upon receiving the states, performs a  measurement in the computational basis. She receives the same post-measurement state as she had sent with a probability of $p=\Big(\sum_{j=0}^3\norm{\ket{E_{0j}}}^2\cdot\norm{\ket{F_{j0}}}^2\Big)$. 
 The term corresponding to $j=0$ is due to a complete correlation between the two parties, however, in the rest of the terms (corresponding to $j=1, 2, 3$), the outcomes of Alice and Bob$_1$ do not match, reflecting the presence of an eavesdropper.

 A similar analysis holds when Alice sends the states $\ket{1}/\ket{2}/\ket{3}$ to Bob$_1$.
 Thus, Eve's interventions introduce errors whenever she gains some information about the key and hence her actions get detected. 

\subsection*{Case II: Bob$_1$ chooses Reflect} In these rounds, in the absence of an eavesdropper, Alice should get the same state as she had sent. Any deviation from this reflects the presence of an eavesdropper. Alice may send states either  from the computational or from the  conjugate basis. We discuss both cases separately.
 
\subsection*{(i) Alice sends a state from the  computational basis}Consider a round in which Alice sends a state $\ket{i},~i\in\{0,1,2,3\}$ to Bob$_1$. The combined state of the traveling system and that of Eve's two ancillae, after the action of two unitaries, can be expressed as:
\begin{align}    
&U_BU_F\ket{i}_T\ket{0}_{F}\ket{0}_{B}= \sum_{j,k=0}^3\ket{k}_T\ket{E_{ij}}_{F}\ket{F_{jk}}_{B}.
\end{align}
Alice gets the same post-measurement state as she had sent with a probability, $p=\norm{\sum_{j=0}^3\ket{E_{ij}}\ket{F_{ji}}}^2$.  Thus, the presence of Eve is detected with a probability $(1-p)$.

\subsection*{\textbf{(ii) Alice sends a state from the  conjugate basis}} Alice may send any one of the states $\ket{j'}$. In that case, the combined state of traveling ququart and those of Eve's two ancillae can be expressed as:
\begin{align}
&\ket{j'}_T\ket{0}_{F}\ket{0}_{B}= \dfrac{1}{\sqrt{4}}\sum_{n,x,y=0}^3e^{\frac{2\pi i j n}{4}}\ket{y}_T\ket{E_{nx}}_{F}\ket{F_{xy}}
_{B}.
\end{align}
From the above equation, it is clear that Alice gets the same post-measurement state as she had sent with a probability $p=\norm{\frac{1}{\sqrt{4}}\sum_{n,x,y=0}^3e^{\frac{2\pi i j (n-y)}{4}}\ket{E_{nx}}\ket{F_{xy}}}^2$. Thus, Eve gets detected with a probability of $(1-p)$.
 This analysis suggests that Eve's interventions introduce errors and hence she gets detected. 
 
 The security of this protocol against attacks due to imperfections such as photon-number-splitting attacks and Trojan-horse attack follows in a similar manner as  discussed in sections (\ref{photon}) and (\ref{Trojan}) for layered QKD protocols.

 The information-theoretic security of the protocol can be performed, following the same approach as adopted for layered QKD in section (\ref{Information-theoretic_security_LQKD}), by combining the information-theoretic security analyses of SQKD for the individual layers, which have already been done in \cite{krawec2015security}.


 \section{Scaling of key rates with dimensions}
 \label{Tradeoff}
 Since dimensionalities of the states employed in key distribution protocols have an impact on key rates \cite{islam2017provably}, it becomes of pivotal importance to study how local dimensionality impacts key rates in a given layer.  By local dimensionality, we mean  the dimensionality of the subsystems belonging to individual participants.

To identify resource states that may give  a higher key generation rate in a specific layer in contrast to other layers, we get back to the procedure given in section (\ref{generalisation}). The procedure starts with the identification of reference multiqubit states. In this case, to have a higher key rate in a specific layer, one needs to employ higher-dimensional reference states instead of qubits. In this section, we show through various examples that  a key rate of $\log_2{r}$ bits in a specific layer can be achieved by employing reference states belonging to the $2^r$-dimensional Hilbert space\footnote{It is assumed that states are sent with an equal probability.}. That is to say, the key rate scales logarithmically with the dimension of the reference state. 
Thus, a specific key rate in a given layer can be achieved by choosing reference states for that layer of appropriate dimensionality.

 To understand this, we consider the same network as considered in the illustrative protocol in section (\ref{Illustrative_QKD}) (which we describe here for sake of completeness).  Note that the  states considered below in the set $S_2$ of two subsystems are Fourier transforms of the states of  the computational bases of respective subsystems. \\
\noindent{\textbf{Network:}} three participants and two layers $L_1$(Alice, Bob$_1$), and layer $L_2$ (Alice, Bob$_1$, and Bob$_2$).\\

\noindent{\textbf{Case I: Resource states with local dimensionalities $(4,2)$ :}}\\

This case has been explored for illustrative protocols with the two sets,
\begin{align}
    &S_1^{(4,2)}: \equiv \{\ket{00}, \ket{11}, \ket{20}, \ket{31}\},~~~
S_2^{(4,2)}\equiv \{\ket{0'+}, \ket{1'-}, \ket{2'+}, \ket{3'-}\}.
\end{align}
In this case, the sifted key rate in both layers is $1$ bit.\\

\noindent{\textbf{Case II: Resource states with local dimensionalities $(3,2)$ :}}\\ 

In this case, the following sets of states, 
\begin{align}
    &S_1^{(3,2)}\equiv\{\ket{00}, \ket{11}, \ket{21}\},~~~S_2^{(3,2)}: \{\ket{0'+}, \ket{1'-}, \ket{2'-}\},
\end{align}
can be employed to implement the task. However, for complete confidentiality of keys between two layers, the rule given in the table (\ref{tab:key_rule}) needs to be followed.\\
\begin{table}[h!]
 \begin{center}
 \resizebox{!}{0.85cm} {
\begin{tabular}{ |c|c|c|c| } 
 \hline
\multirow{2}{2cm}{Bob$_1$'s outcome} & \multirow{2}{2cm}{Bob$_2$'s outcome} & \multirow{2}{2cm}{key symbol in Layer $L_1$}&\multirow{2}{2cm}{ key symbol in Layer $L_2$ }\\ 
&&&\\
 \hline
 $0$ & $0$ & $-$&$0$ \\ 
  \hline
  $1$& $1$ & $1$&$1$\\ 
 \hline
  $2$& $1$ & $0$&$1$\\ 
 \hline
\end{tabular}}
\caption{Key generation rule}
\label{tab:key_rule}
\end{center}
\end{table}
Since Alice sends one of the states from the two sets with an equal probability, the key in layer $L_2$ consists of a larger number of $1's$ than $0's$, thereby reducing the key rate. Similarly, in layer $L_1$, states $\ket{00}$ and $\ket{0'+}$ do not correspond to key generation and are simply discarded. Thus, this choice of resource states leads  to a less random key in layer $L_2$ and does not utilize all the states for key generation in $L_1$. \\

\noindent{\textbf{Case III: Resource states with local dimensionalities $(6,2)$:}}\\ 

In this case, the following sets of states may be employed,
\begin{align}\label{eq:(6.2)}
      &  S_1^{(6,2)}:\{\ket{00},\ket{11},\ket{20},\ket{31},\ket{40},\ket{51}\},\nonumber\\ & S_2^{(6,2)}:\{\ket{0'+},\ket{1'-},\ket{2'+},\ket{3'-},\ket{4'+},\ket{5'-}\},
\end{align}
 to implement the task. Recall that $\ket{j'}=\frac{1}{\sqrt{5}}\sum_{k=0}^5e^{2\pi i j k/6}\ket{k}$.\\
\begin{table}[h!]
 \centering
 \resizebox{!}{1.1cm}{
\begin{tabular}{ |c|c|c|c|c| } 
 \hline
  Bob$_1$ & Key letter generation  & Bob$_2$& layer $L_1$&layer $L_2$  \\ \hline
  0&$2^1\times 0+2^0\times 0\equiv (00)$ & $0$&$0$&$0$ \\ \hline
  1&$2^1\times 0+2^0\times 1\equiv (01)$ & $1$&$0$&$1$ \\ 
 \hline
 2&$2^1\times 1+2^0\times 0\equiv (10)$ & $0$&$1$&$0$ \\ \hline

 3&$2^1\times 1+2^0\times 1\equiv (11)$ & $1$&$1$&$1$ \\ 
 \hline
  4&$2^1\times 2+2^0\times 0\equiv (20)$ & $0$&$2$&$0$ \\ 
 \hline
   5&$2^1\times 2+2^0\times 1\equiv (21)$ & $1$&$2$&$1$ \\ 
 \hline
\end{tabular}}
\caption{Key generation rule for states with local dimensionality $(6,2)$.}
\label{tab:key_rule_3}
\end{table}
As per  the key generation rule given in the table (\ref{tab:key_rule_3}), key symbols in both layers are generated with equal probabilities. So, sifted key rates in layers $L_1$ and $L_2$ are $(\log_2{3})$ bit and $1$ bit respectively.

Thus, a key rate of $\log_2r$ bits can be achieved by choosing reference states belonging to $2^r$-dimensional Hilbert space as reflected in cases I and III. In case II, we have presented an example that allows sharing simultaneous keys in both layers. However, it does not utilise all the states for key generation in both layers.
 These examples illustrate that different key rates can be achieved by employing resource states with different local dimensionalities.  We have given an illustration to identify resource states for case III given in equation (\ref{eq:(6.2)}) in Appendix (\ref{procedure_dimension}).


\section{Conclusion}\label{conclusion}
In summary, this work explores the potential offered by multidimensional separable states in both QKD and SQKD protocols in layered networks. We have shown the robustness of the protocols  against various eavesdropping strategies. The probability of detection of eavesdropping grows exponentially with the number of rounds. The protocols may be made resilient against photon-number-splitting attacks by incorporating decoy pulses. We have shown that pinpointing the location of Eve helps in identifying layers in which keys can be shared securely even if there is eavesdropping in others. We have studied the scalability of key rates with local dimensionality of states that allows achieving a specific key rate in a given layer.
 Employment of multidimensional separable states makes (i) implementation of these protocols more feasible and that too (ii) with a greater yield than entanglement-based protocols.    
 
 This work opens up many avenues for further study of quantum communication protocols in networks. We wish to stress that we have shown only QKD and SQKD protocols in layered networks. However, in general, different kinds of secure quantum communication protocols, e.g., QKD and quantum secure direct communication protocols may also be clubbed together.   That forms an interesting study.

\begin{appendices}
\section{A brief recapitulation of SQKD protocol \cite{Boyer07}}\label{SQKD_Boyer}
The steps of the protocol proposed in \cite{Boyer07} are as follows:
\begin{enumerate}
    \item Alice randomly prepares one of the states $\{\ket{0}, \ket{1}, \ket{+}, \ket{-}\}$ with an equal probability and sends it to Bob. 
    \item Bob, upon receiving the state, exercises one of the two options with equal probabilities: (i) he measures the incoming state in the computational basis. He prepares the same state as the post-measurement state afresh and sends it back to Alice, (ii) He sends the incoming state back to Alice.  
    \item Alice measures each incoming state in the same basis in which she has prepared it. This constitutes one round.
    \item After a sufficient number of rounds, Bob reveals the rounds in which he has performed measurements  and Alice reveals the rounds in which she has sent the states in the computational basis, i.e., $\{\ket{0}, \ket{1}\}$. 
    \item Alice analyses the data of those rounds in which Bob has not performed measurements to check for the presence of an eavesdropper.
    \item The outcomes of those rounds, in which Bob has performed measurements and Alice has measured in the computational basis, constitute a key.
\end{enumerate}
In this way, a key is shared between Alice and Bob.
\section{Illustration for identification of resource states}\label{procedure_dimension}
The sets of resource states, given in equation (\ref{eq:(6.2)}), can be identified by choosing reference states of qubits for distributing the key in layer $L_2$ and reference states of qutrits for distributing the key in layer $L_1$. These sets are given by,
\begin{align}\label{eq:resource}
  &   {\cal S}_1^{(1)}:\{\ket{0},\ket{1},\ket{2}\},~~~~{\cal S}_2^{(1)}:\{\ket{0'},\ket{1'},\ket{2'}\},\nonumber\\
    & {\cal S}_1^{(2)}:\{\ket{00},\ket{11}\},~~~~~{\cal S}_2^{(2)}:\{\ket{++},\ket{--}\}.
 \end{align}
 In layer $L_1$, there are only two participants, {\it viz.}, Alice and Bob$_1$. It is sufficient for Alice to send monoparty states to Bob$_1$ in the layer $L_1$. However, in layer $L_2$, Alice needs to send states to both Bob$_1$ and Bob$_2$, so sets ${\cal S}_1^{(2)}$ and ${\cal S}_2^{(2)}$ have bipartite states. Following the prescription of section (\ref{generalisation}) with the resource states given in equation (\ref{eq:resource}), the two sets that distribute keys in the network are,
 \begin{align}
 & {\cal S}_1:\{\ket{000},\ket{011}\ket{100},\ket{111}\ket{200},\ket{211}\},\nonumber\\
 &{\cal S}_2:\{\ket{0'++},\ket{0'--},\ket{1'++},\ket{1'--},\ket{2'++},\ket{2'--}\}.
 \end{align}
 The first two subsystems belong to Bob$_1$ while the third one belongs to Bob$_2$. Employing binary-to-decimal mapping, bipartite states of qutrit and qubit can be mapped to single-party six-dimensional states as follow:
 \begin{align}
 & S_1:\{\ket{00},\ket{11}\ket{20},\ket{31}\ket{40},\ket{51}\},\nonumber\\
 &S_2:\{\ket{0'+},\ket{1'-},\ket{2'+},\ket{3'-},\ket{4'+},\ket{5'-}\}.
 \end{align}
 The sets $S_1$ and $S_2$ are the same as given in equation (\ref{eq:(6.2)}) of section (\ref{Tradeoff}).
 
\end{appendices}

\section*{Acknowledgement}
We thank the anonymous referee for the valuable comments and suggestions which have brought more clarity to the manuscript. Rajni thanks UGC for funding her research. Sooryansh thanks CSIR (Grant no.: 09/086 (2017)-EMR-I) for funding his research.
\large

\section*{Data availability statement}
Data sharing is not applicable to this article as no datasets were generated or analyzed during the current study.
\section*{Disclosures}
The authors declare no conflicts of interest.
\section*{Author contribution statement}
All the authors have contributed equally in all respects.
\end{document}